**Why So Many Published Sensitivity Analyses Are False.**

**A Systematic Review of Sensitivity Analysis Practices**

Andrea Saltelli (1), Ksenia Aleksankina (2), William Becker (3), Pamela Fennell (4),

Federico Ferretti (3), Niels Holst (5), Li Sushan (6), Qiongli Wu (7)

(1) University of Bergen, NO & Universitat Oberta de Catalunya, ES

(2) University of Edinburgh & NERC Centre for Ecology and Hydrology, Edinburgh, UK

(3) European Commission, Joint Research Centre, Ispra, IT

(4) University College of London, UK

(5) Aarhus University, DK

(6) Technische Universität Darmstadt, Darmstadt, DE

(7) Wuhan Institute of Physics and Mathematics, Chinese Academy of Sciences, Wuhan, CHINA

**Abstract**

Sensitivity analysis (SA) has much to offer for a very large class of applications, such as model selection, calibration, optimization, quality assurance and many others. Sensitivity analysis provides crucial contextual information regarding a prediction by answering the question 'Which uncertain input factors are responsible for the uncertainty in the prediction?' SA is distinct from uncertainty analysis (UA), which instead addresses the question 'How uncertain is the prediction?' As we discuss in the present paper much confusion exists in the use of these terms.

A proper uncertainty analysis of the output of a mathematical model needs to map what the model does when the input factors are left free to vary over their range of existence. A fortiori, this is true of a sensitivity analysis. Despite this, most UA and SA still explore the input space; moving along mono-dimensional corridors which leave the space of variation of the input factors mostly unscathed.

We use results from a bibliometric analysis to show that many published SA fail the elementary requirement to properly explore the space of the input factors. The results, while discipline-dependent, point to a worrying lack of standards and of recognized good practices.

The misuse of sensitivity analysis in mathematical modelling is at least as serious as the misuse of the p-test in statistical modelling. Mature methods have existed for about two decades to produce a defensible sensitivity analysis. We end by offering a rough guide for proper use of the methods.

**Introduction**

**1. Background**

*1.1 The problem setting and study objectives*

With the increase of computing power in recent decades, mathematical models have become increasingly prominent tools in decision-making processes in engineering, science, economics and policy-making, among other applications. Coupled with the abundance of available data, models have also become increasingly complex—examples include large climate or economic models, which aim to include ever more processes at an ever-higher resolution. However, this increased complexity requires much more information to be specified as model inputs, and typically this information is not well-known. It is therefore essential to understand the impact of these uncertainties on the model output, if the model is to be used effectively and responsibly in any decision-making process.

Sensitivity analysis is "[t]he study of how the uncertainty in the output of a model (numerical or otherwise) can be apportioned to different sources of uncertainty in the model input" [1]. As such it is very much related to – but distinct from – uncertainty analysis (UA), which instead characterizes the uncertainty in model prediction. Ideally an uncertainty analysis precedes a sensitivity analysis: before uncertainty can be apportioned it needs to be estimated. However, this consideration is not universally shared; while most practitioners of SA distinguish it from UA, modellers overall tend to conflate the two terms. We shall discuss in a moment why this is the case.

Sensitivity analysis is used for many purposes. Primarily it is used as a tool to quantify the contributions of model inputs, or groups of input, to the *uncertainty* in the model output— examples of such applications include Eisenhower et al.[2] and Becker et al.[3]. This use of sensitivity analysis will be the focus of the present paper. In this uncertainty setting, typical objectives are to identify which input factors contribute the most to model uncertainty ("factor prioritisation") so that further information might be collected about these parameters to reduce model uncertainty, or to identify factors which contribute very little and can potentially be fixed ("factor fixing") [4].

In engineering design, "design sensitivity analysis" is used as a tool for structural optimisation [5]. Sensitivity analysis can also be used to better understand processes within models, and thereby, the natural systems on which they are based [6], or as a quality assurance tool: an unexpected strong dependence of the output upon an input deemed irrelevant might either illuminate the analyst on an unexpected feature of the system or reveal a conceptual or coding error. Desirable properties of sensitivity analysis strategies are discussed in Saltelli [1], see Box 1.

This paper has the following objectives:

- To assess the "state" of sensitivity analysis across a range of academic disciplines. We do this by a systematic review of a large number of highly-cited papers in which sensitivity analysis is the focus in some respect.
- To discuss – based on this review - known problems and misinterpretations of sensitivity analysis, and propose some ideas for how these problems might be addressed.

> **Box 1.** Desirable properties of a sensitivity analysis methods (from Saltelli)[1]. Variance based measures[9] and moment independent methods[34] are two classes of sensitivity measures which possess these properties.
>
> 1. The ability to cope with the influence of scale and shape. The influence of the input should incorporate the effect of the range of input variation and the form of its probability density function (pdf). It matters whether the pdf of an input factor is uniform or normal, and what the distribution parameters are.
>
> 2. To include multidimensional averaging. In an OAT (for moving One factor At a Time) approach to SA, e.g. using partial derivatives $S_j = \partial Y/\partial X_j$, one computes the effect of the variation of a factor when all others are kept constant at the central (nominal) value. A global method should instead evaluate the effect of a factor while all others are also varying.
>
> 3. Being model independent. The method should work regardless of the additivity or linearity of the model. A global sensitivity measure must be able to appreciate the so-called interaction effect, which is especially important for non-linear, non-additive models. These arise when the effect of changing two factors is different from the sum of their individual effects.
>
> 4. Being able to treat grouped factors as if they were single factors. This property of synthesis is essential for the agility of the interpretation of the results. One would not want to be confronted with an SA made of large tables of sensitivity measures.

To set the scene let us describe UA first: how is it done? If all the uncertainty in the model output comes only from the model input factors – i.e. if the model does not generate additional uncertainty (deterministic models), then assessing the uncertainty in the output boils down to propagating the uncertainty from the input factors to the output, for example by repeatedly running the model using different values for the uncertain inputs within their plausible ranges. This could be done with a Monte Carlo simulation, or with some ad hoc design, to generate a distribution of possible model results (the grey area in Figure 1).

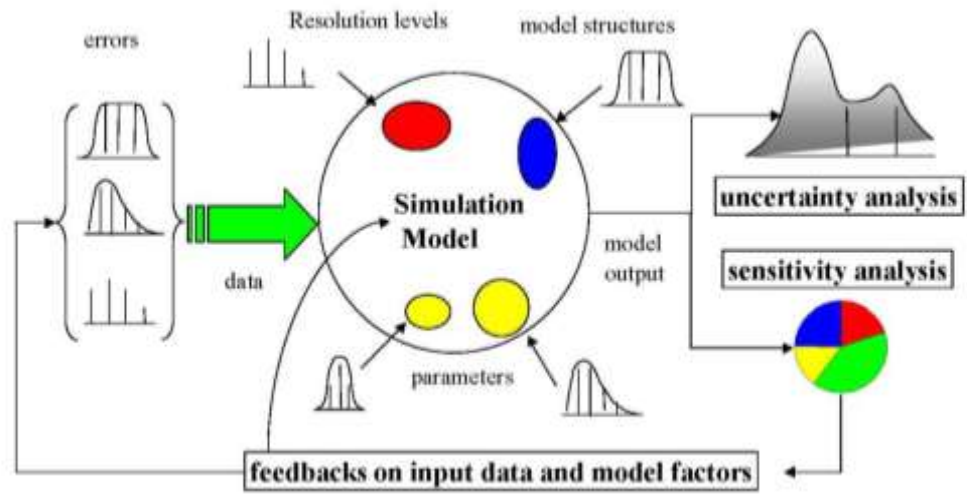

*Figure 1, Idealized uncertainty and sensitivity analysis. Uncertainty coming from heterogeneous sources is propagated through the model to generate an empirical distribution of the output of interest (grey curve). The uncertainty in the model output, captured e.g. by its variance, is then decomposed according to source, thus producing a sensitivity analysis.*

Characterizing such a distribution – e.g. by constructing it empirically from the output data points, constitutes an uncertainty analysis. The UA would then be completed by extracting summary statistics, such as the mean, median, and variance, from this distribution and possibly by assigning confidence bounds, e.g. on the mean.

Once this is done the next step could be to distribute this uncertainty among the input factors, to achieve inference of the type "Factor $x_i$ alone is responsible for 70% of the variation in the output". This type of inference is what we would call a sensitivity analysis.

In sensitivity analysis and experimental design, it is often helpful to think of the set of all possible combinations of input factors as an "input space". For example, with two model inputs, any combination of values could be marked as a point on a two-dimensional plane, with the range of factor 1 on one axis, and the range of factor 2 on the other. In the case of three input factors the input space would be a cube, and for higher numbers, a hypercube. A key message of this paper is that an important fraction of the uncertainty and sensitivity analyses seen in the literature leave unexplored a very large fraction – if not almost the totality – of the input space. As a consequence, these analyses don't propagate the uncertainty from the input to the output, but only a small – at times minuscule – part of it, i.e. they fail the preliminary stage of estimating the uncertainty – see Section 1.4 for a geometric interpretation of this statement.

How can this error happen? There is here a hierarchy of causes which we could order as follows:

- Modelling includes elements of craft as much as science[7]; as such every discipline goes about modelling following local disciplinary standards and practices;
- If modelling is a non-standardised discipline the same holds a fortiori for uncertainty and sensitivity analysis, hence the difficulty for good practices to establish themselves. Researchers from different fields have difficulties to communicate with one another in a transversal discipline;
- Most scientists conflate the meaning of SA and UA. In a large class of instances (e.g. in economics) SA is understood as an analysis of the robustness of the prediction (UA);
- Most modellers prefer to change factors one at a time as a result of their training and methodological disposition to think in terms of derivatives;
- As seen with the 'reproducibility crisis' in scientific work, some researchers simply don't have enough knowledge and training in statistics;
- Although mature global sensitivity analysis methods have been around for more than 25 years, this still may not be enough time for established good practice to filter down into the many research fields in which modelling is used.

Before turning to our literature review, we shall briefly touch upon these problems in turn.

*1.2 If modelling as a craft, what is then sensitivity analysis?*

Robert Rosen, a system ecologist, tackles the specificities of modelling in the scientific method in his work 'Life Itself' [7]. Here he suggests that when a model is built to represent a natural system we should look at the play of causality. The argument is that the natural system is kept together – Rosen uses the word 'entailed' - by *material*, *efficient* and *final* causality. In contrast, the formal system, i.e. the model, is only internally entailed by *formal* causality. Rosen uses here the four causality categories of Aristotle, on which we will not dwell here, to highlight that no arrow of causality flows from the natural system to the formal one. In other words, the act of encoding (Figure 2) is not driven by causality, which would fix the model specification, but is driven by the needs and the craft of the modeller. Otherwise said, different modelling teams given the same data can produce altogether different models and inference [8].

Thus, the success of the modelling operation is judged by the usefulness – or otherwise - of the insights made possible by the operation of decoding, which is another way of saying that all models are wrong but some are useful – according to an aphorism attributed to George Box.

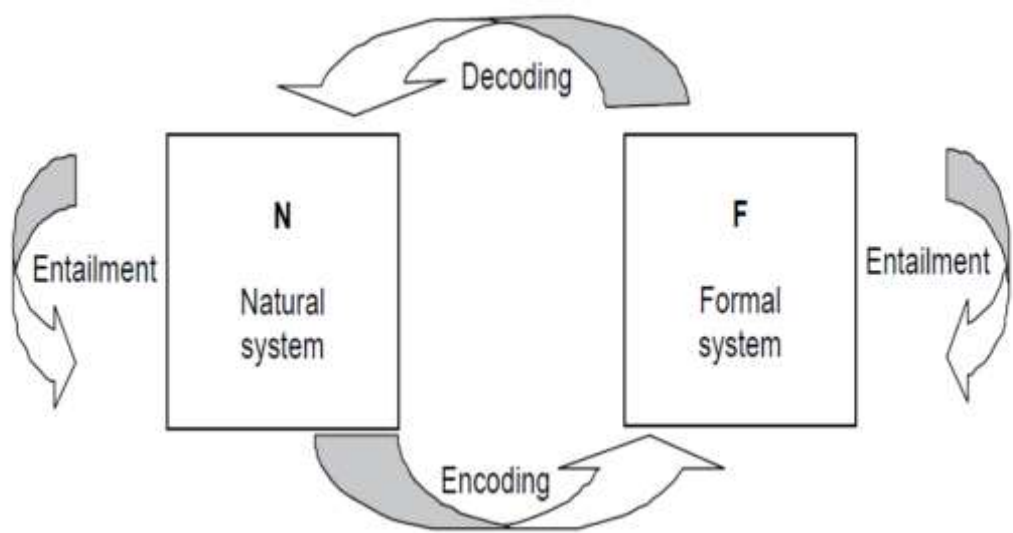

*Figure 2: The modelling relation following Rosen (1991). For a discussion see Saltelli et al. [9].*

An independent take on the specificity of modelling is due to Naomi Oreskes [10], a historian. She takes issue with the ill-considered use of modelling in prediction, by observing that model-based predictions tend to be treated as logical predictions as seen in the hypothetic-deductive model of science. In this scheme of how science operates[11] a hypothesis (e.g. a hypothesized physical law) is linked to a deduction - what should happen if the hypothesis were true, and, when the expected deduction does not take place experimentally, one says that the hypothesis has been 'falsified', in the sense of proven false.

For Oreskes, to be of value in theory testing, the model-based predictions "must be capable of refuting the theory that generated them". But models are not crisp physical laws, they are "complex amalgam of theoretical and phenomenological laws (and the governing equations and algorithms that represent them), empirical input parameters, and a model conceptualization." Thus, when a model prediction fails what part of all this construct was falsified? The input data? The system's formalization? The algorithms used in the model?

This short discussion - based on our arbitrary choice of sources and glossing over half a century of disputes over the nature of the scientific method, serves to illustrate why modelling is so discipline-specific (see in this respect a recent review by Padilla et al.[12], and why - as a result, even relatively straightforward methodologies which are ancillary to modelling, such as uncertainty and sensitivity analysis, are not part of a standardized syllabus being taught across disciplines. We shall see in our review below that this is reflected in a very heterogeneous and often unsatisfactory approach to UA, and SA, both across and within disciplines.

*1.3 Uncertainty or sensitivity?*

Many practitioners accept a taxonomy of sensitivity analysis based on distinguishing between local and global methods [9]. A local method in its simplest form is the derivative of a scalar output $y$ – the output of the model in this case - with respect to one of its input factor $x_i$. Thus, a mathematical sensitivity is just $s_i = \frac{dy}{dx_i}$. A global sensitivity analysis method, at the other extreme, could be an analysis of variance (ANOVA) as usually taught in experimental design, which informs the analyst about factors' influence in terms of its contribution to the variance of the model output, including the effect of interaction among factors [13]. Another example of a global measure is Pearson's correlation ratio

$$\eta^2 = \frac{V_{x_i}\left(E_{x_{\sim i}}(y|x_i)\right)}{V(y)}$$

where $V(y)$ is the unconditional variance of $y$, obtained when all factors $x_i$ are allowed to vary, and $E_{x_{\sim i}}(y|x_i)$ is the mean of $y$ when one factor is fixed. In sensitivity analysis, $\eta^2$ is also known as a first order sensitivity index. This is a statistical measure of sensitivity in a class which is termed 'variance-based'. Its meaning can be expressed in plain English: $\eta^2$ is the expected fractional reduction in the variance of $x_i$ that would be achieved if factor $x_i$ could be fixed. $\eta^2 = 1$ implies that all variance of $y$ is driven by $x_i$, and hence that fixing it also determines *y*.

A further discussion of this and of the theory of sensitivity indices is beyond the scope of this paper and the reader is referred e.g. to Saltelli et al.[9].

Confusingly, the term 'sensitivity analysis' is also used in economics to mean in fact an uncertainty analysis. This is perhaps due to an influential paper of econometrician Edward Leamer [14], entitled "Sensitivity analysis would help", whose problem setting and motivation were to ensure the robustness of a regression analysis with respect to various modelling choices, e.g. in the selection of regressors. As a result, in economics and finance it is common to see the expression 'sensitivity analysis' used to mean what we have defined here as uncertainty analysis.

*1.4 Pitfalls and charm of one-factor-at-a time*

One very common way that sensitivity analysis is performed in practice is by moving one factor at a time (OAT). In short, this involves keeping all input factors at nominal values, then moving one input to its maximum and minimum values, and observing the effect on the output. This process is repeated for all input factors, always keeping all other input factors fixed except the one that is being perturbed. Figure 3 (left) illustrates an OAT design with two input factors, and a corresponding global design (right) that might be used to estimate the global measures discussed in the previous section.

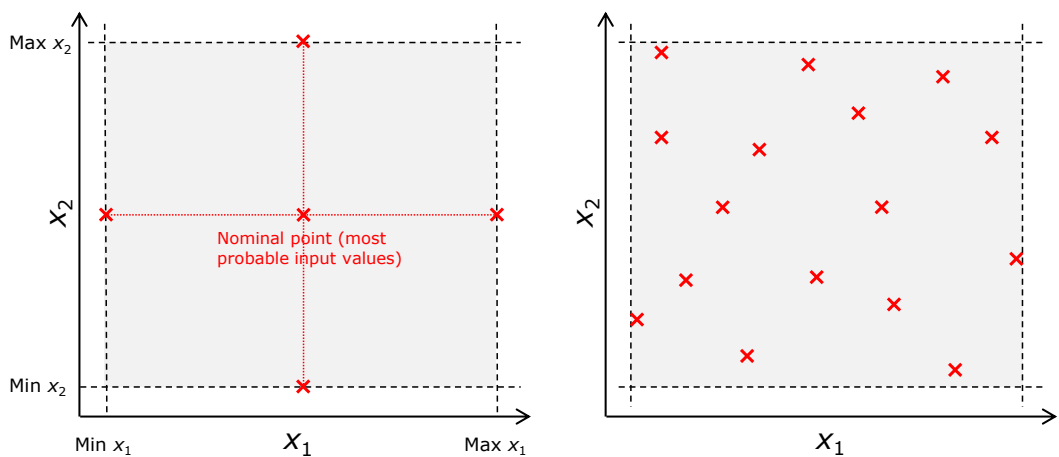

*Figure 3: OAT design (left) contrasted against global design (right)*

As detailed in a previous paper [15], and evident from 4, this is a very restrictive way to explore a multidimensional space. We can illustrate this with a simple example. Imagine that the input space is a three-dimensional cube of side one. Moving one factor at a time by a distance of ½ away from the centre of the cube generates points on the faces of the cube, but never on its corners. All these points are in fact on the surface of a sphere internal and tangent to the cube, as illustrated in Figure 4. The area of the sphere divided the area of the cube is about ½. So far, we do not seem to have achieved much, other than to say that there is about one half of the area of the cube with no points. In fact, if we increase the number of dimensions this ratio goes towards zero very quickly. In ten dimensions, the area of the hypersphere divided by the area of the hypercube is 0.0025, one fourth of one percent. In practice, it is even more restrictive than that because the OAT design does not even fully explore inside the hypersphere, and is limited to a "hypercross". In other words, moving factors OAT in ten dimensions leaves over 99.75% of the input space unexplored. This under-exploration of the input space directly translates into a deficient sensitivity analysis, and is but one of the many incarnations of the so-called "curse of dimensionality", and the reason why an OAT SA is perfunctory, unless the model is proven to be linear.

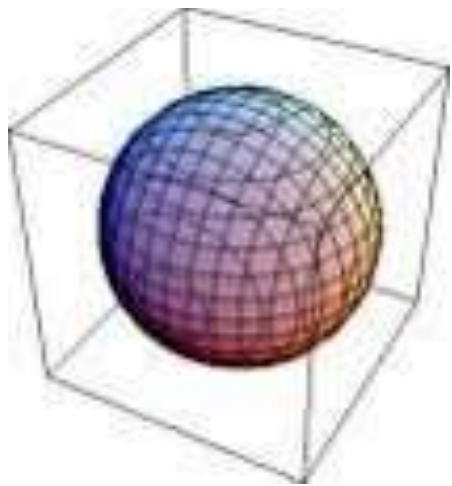

*Figure 4: A sphere in included in a cube (three-dimensional case) and tangent to its faces. The volume of the sphere divided that of the cube is roughly 1/2. If the dimension were ten instead of three the same ratio would be 0.0025.*

Statisticians are well acquainted with this problem. This is why, in the theory of experimental design[13] factors are moved in groups, rather than OAT, to optimize the exploration of the space of the factors.

Why then, do modellers persistently revert to OAT? A trivial explanation would be that they are unaware of experimental design. There are in fact a host of other reasons: for one, the more factors we move, the higher the chance that the model will crash or misbehave. Note that this is precisely the reason why a global SA is a good instrument of model verification: it is unusual to run a global SA without detecting model errors – modellers call this jokingly

Lubarsky's Law of Cybernetic Entomology, according to which 'there is always one more bug'.

Next, modellers often feel comfortable about the baseline point (the centre of the cube in the previous example) and do not wish to depart too much from it. Then comes the ease of interpretation: if I move just one factor $x_i$ then the change I observe in $y$ must come from $x_i$ alone. Finally, as noted by the same Leamer in a subsequent paper [16], the reluctance to take up these methods may be due to their candour.

A proper method, by honestly propagating all of the input uncertainty, may lead to an inconveniently wide distribution of the output of interest. For example, a cost benefit analysis reporting a distribution encompassing possible large losses as well as large gains may not be what the owner of the problem wishes to hear. This is the same as to say that the volatility of the inference is exposed, and thus is the insufficiency of the evidence. According to Leamer [16], as well as to [17], this situation may induce modellers to 'massage' the uncertainty in the input factors so that the output falls in a more desirable zone.

For this reason, when performing sensitivity auditing (an extension of sensitivity analysis for policy-loaded cases, Saltelli et al.[18]) the analyst is required to perform a check on possible inflation or deflation of the uncertainties in the input factors. These elements will resurface in our discussion of the p-value later in the text.

*1.5 What do we define as 'OAT' and what do we take as 'global'?*

The identification of OAT and global sensitivity analyses is one of the focal points of this meta-study. In the literature review presented in this work we have defined OAT methods as all derivative based approaches where factors are moved only one at a time, even when derivatives are computed efficiently, such as when using the adjoint method [19]. We have also called OAT approaches where the factors are moved by a finite step $\Delta x_i$ – provided that only one factor is moved. Note that some methods, such as that of Kucherenko[20] or Morris[21] *are* based on derivatives but are classified as global methods because they sample partial derivatives or incremental ratios at multiple locations in the input space.

We have defined as global any approach that is based on moving factors together, such as in Design of Experiment (DoE). A Monte Carlo analysis followed by an analysis of the scatterplots of $y$ versus the various input factors $x_i$ is also classified as global, as well as approaches based on regression coefficients of $y$ versus the $x_i$s, the use of Sobol' sensitivity indices - independently of the way these are computed, screening methods such as the method of Morris, Monte Carlo filtering, various methods known as 'moment-independent' and so on, see Saltelli et al.[9] for a description, and the additional online material for the methods met in the papers reviewed.

*1.6 Is sensitivity analysis a discipline? Who decides if a method is a good practice?*

In defining all global methods as a recommended good practice, the authors of the present work are taking upon themselves the responsibility of a speaking for a community of practitioners. One such community might be said to have formed around a series of SAMO conferences (for (sensitivity analysis of model output, see http://samo2016.univ-reunion.fr/). SAMO is held every three years since 1995. This community is active in disseminating good practices, e.g. via entries devoted to sensitivity analysis in encyclopaedias, see e.g. Iooss and Saltelli [22], Becker and Saltelli [22], for recent examples. Yet a caveat is in order as to the legitimacy of the authors' claim.

First, the authors cannot speak for the entire SAMO community. Next, SAMO does not capture the full spectrum of practitioners interested in uncertainty and sensitivity analysis. For example, in the United States, SA-related activities are under the heading of 'Verification, Validation and Uncertainty Quantification' (VVUQ), for which a journal of the American Society of Mechanical Engineers is available (http://verification.asmedigitalcollection.asme.org/journal.aspx).

The importance of sensitivity analysis is generally acknowledged. Sensitivity analysis is prescribed in national and international guidelines in the context of impact assessment (e.g. European Commission, p. 390-393 [23]; US Office for the Management and Budget [a]; US Environmental Protection Agency[b]). When the output of a model feeds into policy prescription and planning, a sensitivity analysis would appear as an essential element of due diligence.

In conclusion, while some communities do practice sensitivity analysis, these studies are not part of a recognized syllabus nor does sensitivity analysis constitute a discipline. This explains the heterogeneous natures of the results discussed later in this work.

*1.7 Analogies and differences with the use and misuse of the p-value*

It is perhaps instructive to situate the present work in the context of a general reflection on the quality of scientific work, and the role of statistics therein. A paper published in 2005 by John Ioannidis [24] warned about the poor quality of most published research results. The paper was taken up by the media, and the periodical "The Economist" devoted its cover to the issue in 2013, with a full article describing the subtleties of use and misuse of statistics in deciding about the significance of scientific results. The specific subject of concern was the use of the p-value, a fundamental tool used by researchers to decide if a given result is just the result of chance or indeed an effect worth publishing.

---

[a] OMB, Proposed risk assessment bulletin, Technical report, The Office of Management and Budget's – Office of Information and Regulatory Affairs (OIRA), January 2006, https://www.whitehouse.gov/sites/default/files/omb/assets/omb/inforeg/proposed_risk_assessment_bulletin_010906.pdf, pp. 16–17, accessed December 2015.

[b] EPA, 2009, March. Guidance on the Development, Evaluation, and Application of Environmental Models. Technical Report EPA/100/K-09/003. Office of the Science Advisor, Council for Regulatory Environmental Modeling, http://nepis.epa.gov/Exe/ZyPDF.cgi?Dockey=P1003E4R.PDF, Last accessed December 2015.

In 2016, the pressure surrounding the statistical community was so high that the American Statistical Association felt the need to intervene with a statement[25] to clarify how the test should be used. Useful reading on the topic are Gigerenzer and Marewski[26] and Colquhoun[27]. These articles show a complex mix of causes – from poor training to bad incentives – which result in the generalized failure in the use of the p-value, evidenced by attempts to repeat published results (see e.g. Shanks et al.[28]).

The problem is seen as a combination of confirmation bias - authors looking for the effect they presume will be there, or effect-less papers remaining unpublished, of p-hacking – changing the setup of the study or the composition of the sample till an effect emerges, and HARKing (formulating the research Hypothesis After the Results are Known, Kerr[29]). The latter involves repeatedly running comparison tests between different combinations of variables until a "significant" result is found, without having a prior hypothesis.

As for the case of sensitivity analysis, a proper use of the p-value could identify weak inference, and decreases the chance that a given result is considered worth publishing. For example, many investigators, when using the p-value with a 0.05 (5%) significance level, believe that rejecting the null hypothesis implies only one chance in 20 of being wrong, while when relevant information about the prevalence (expected fraction of effects before the experiment is run) and the power of the test (linked to the number of false negative) are included, then the chance of erroneous identification could turn out to be as high as one in three, or even higher [27].

The equivalent case for uncertainty analysis would be one where – once all factors are allowed to change over their range of existence, the inference were to result 'diffuse', e.g. spread over several orders of magnitudes and hence useless. Funtowicz and Ravetz [17] have defined as pseudo-science, precisely the case "where uncertainties in inputs must be suppressed lest outputs become indeterminate".

It is clear that the p-value is a topic for statisticians, which makes it possible for the disciplinary community to condemn its misuse as just discussed [25], although the discussion within the community was intense and led to as many as twenty dissenting commentaries (see http://amstat.tandfonline.com/doi/suppl/10.1080/00031305.2016.1154108?scroll=top). In the case of sensitivity analysis, one lacks a disciplinary community. Hence the attempt made in the present paper must be considered as just the expression of a concerned group of scholars. Note that the present discussion on the crisis of reproducibility, statistics, and science overall is in full swing. Every day brings new elements of analysis[30], alarm as to the fate of entire sub-disciplines[31] and concerned warnings as to the perverse dynamics of the system[32,33].

## 2. The literature review

In order to understand the prevalence and type of sensitivity analysis across different fields, an extensive literature review (a meta-study) was carried out. The review was based on highly-cited articles that have a focus on sensitivity analysis. The reasoning here was that the most highly cited articles should represent, on average, "good practice" relative to that field. Therefore, by analysing these papers, we should be able to conclude, with reasonable confidence, that the rigour of sensitivity analysis in a given field is at, or below, the level of its top-cited papers.

*2.1 Design of the experiment*

The literature search was conducted on the Scopus database. In order to identify relevant papers, the following search criteria were used (after a few iterations of analysis and refinement)[c] . First, the strings "sensitivity analysis" *and* "model/modelling", *and* "uncertainty" were required to be present in the title, abstract or keywords. This ensures that the paper has a significant focus on sensitivity analysis, that it is related to mathematical

---

[c] Exact query specifications available in the Additional Online Material. Retrieved from https://www.scopus.com between March and May 2017

models, and concerns uncertainty (as opposed to e.g. design sensitivity analysis and optimisation, which is a separate topic). Second, the papers were restricted to the years 2012-2017, in order to provide a sample of recent research. Finally, the results were required to be journal articles, and in English (the latter for ease of reviewing).

This search resulted in around 6000 articles. The search query is deliberately restrictive, in that sensitivity analysis articles exist that do not mention "model" in the abstract, title or keywords, for example. However, it was considered to be an unbiased way of automatically selecting sensitivity analysis papers across fields. Preliminary attempts indicated that simply mentioning "sensitivity analysis" yielded far too many irrelevant articles (around 47,000). The sample here therefore can be considered as representative, but the numbers of papers returned are significantly below the true number of sensitivity analysis papers in the literature.

Each paper returned by the search is tagged using one or more subject identifiers. Subject areas with less than 100 articles meeting the search criteria (of which there were eight) were not examined in this study. The resulting 19 subject areas are as follows:

1. AGR_BIO_SCI (Agricultural and Biological Sciences)
2. BIOCHEM_GEN_MBIO (Biochemistry, Genetics and Molecular Biology)
3. BUS_MAN_ACC (Business, Management and Accounting)
4. CHEMI (Chemistry)
5. CHEM_ENG (Chemical Engineering)
6. COMP_SCI (Computer Science)
7. DEC_SCI (Decisional Science)
8. EARTH_SCI (Earth and Planetary Sciences)
9. ECON_FIN (Economy and Finance)
10. ENERGY (Energy)
11. ENGINEERING (Engineering)
12. ENV_SCI (Environmental Science)
13. IMMUN_MICROBIO (Immunology and Microbiology)

14. MAT_SCI (Mathematical Science)

15. MATHS (Maths)

16. MEDICINE (Medicine)

17. PHAR_TOX (Pharmacology and Toxicology)

18. PHYS_ASTRO (Physics and Astronomy)

19. SOC_SCI (Social Science)

To understand the occurrence of sensitivity analysis across disciplines, Figure 5 shows the distribution of sensitivity analysis papers across research fields, by density (number of SA papers divided by total number in the search period) and by number. The greatest density of papers is found in decision science, as well as model-intensive subjects such as earth sciences, environmental science and energy. The greatest raw numbers are found in environmental science, engineering, and medicine, where the latter does not have a high density due to the very large overall research output. Note that articles can be tagged with more than one subject identifier.

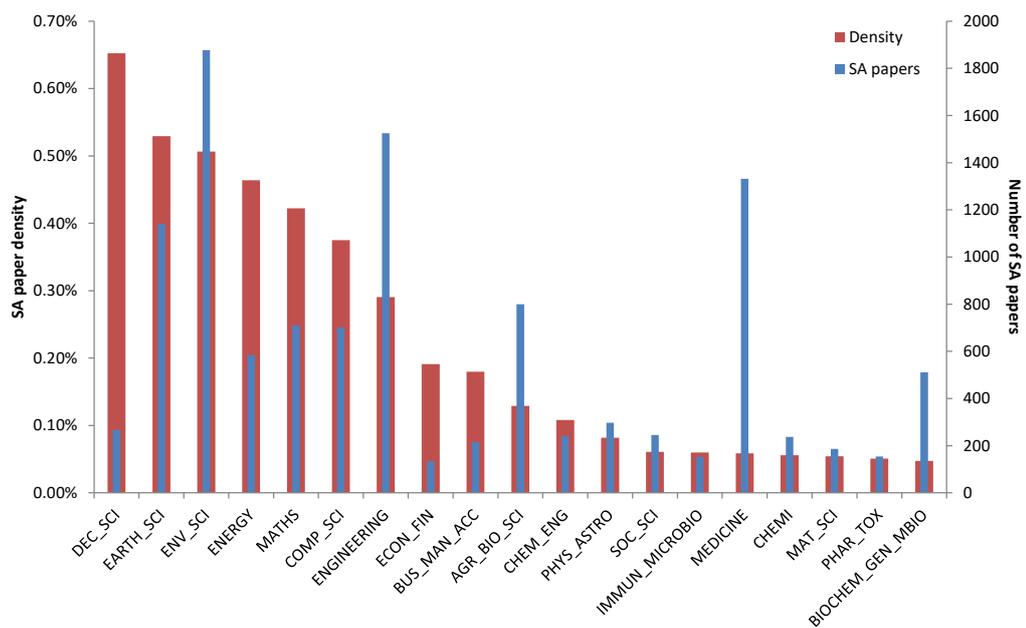

*Figure 5: Density and number of sensitivity analysis articles returned by search criteria, by subject*

In order to provide a manageable sample of articles for review, the top twenty most-cited papers from each field were selected. Since most papers include more than one subject identifier, some papers featured in more than one of the top-twenty lists. The reviewing was distributed between the authors of the present article. Even though the initial search criteria had been refined to focus on model-related sensitivity analysis, a total of 44 papers had to be discarded as not including a sensitivity analysis, nor an uncertainty analysis, or because they reported an analysis of the dependence of the output upon just one factor (which does not constitute a sensitivity analysis). A total of 280 papers were finally retained for the analysis, though in total 324 papers were reviewed.

The scoring matrix as well as the authors' review notes are given in the Additional Online Material, while a summary is presented in Table 1.

## 3. Results

Table 1 shows the review's results in a condensed form.

*Table 1 Summary results*

| Category | METHOD | | | | | MODEL LINEARITY | | | PAPER FOCUS | | Total |
|---|---|---|---|---|---|---|---|---|---|---|---|
| | Global SA | OAT SA | Global UA | OAT UA | Other/Unclear | Linear | Nonlinear | Unclear | Method | Model | reviewed |
| AGR_BIO_SCI | 15 | 11 | 6 | 0 | 6 | 1 | 22 | 4 | 3 | 24 | 27 |
| BIOCHEM_GEN_MBIO | 23 | 15 | 6 | 1 | 7 | 2 | 19 | 15 | 0 | 36 | 36 |
| BUS_MAN_ACC | 4 | 7 | 5 | 5 | 1 | 1 | 18 | 2 | 3 | 18 | 21 |
| CHEMI | 10 | 8 | 2 | 0 | 5 | 0 | 17 | 5 | 1 | 21 | 22 |
| CHEM_ENG | 12 | 12 | 4 | 0 | 5 | 0 | 16 | 12 | 1 | 27 | 28 |
| COMP_SCI | 21 | 9 | 1 | 1 | 2 | 8 | 16 | 6 | 11 | 22 | 33 |
| DEC_SCI | 9 | 7 | 3 | 4 | 0 | 2 | 20 | 1 | 7 | 15 | 22 |
| EARTH_SCI | 11 | 13 | 4 | 1 | 17 | 5 | 13 | 24 | 2 | 41 | 43 |
| ECON_FIN | 5 | 8 | 6 | 3 | 0 | 1 | 16 | 1 | 0 | 18 | 18 |
| ENERGY | 14 | 15 | 3 | 4 | 2 | 3 | 17 | 16 | 0 | 36 | 36 |
| ENGINEERING | 38 | 16 | 5 | 5 | 5 | 3 | 51 | 11 | 3 | 62 | 65 |
| ENV_SCI | 31 | 22 | 14 | 4 | 16 | 6 | 44 | 24 | 11 | 67 | 78 |
| IMMUN_MICROBIO | 19 | 7 | 3 | 0 | 5 | 2 | 6 | 13 | 0 | 21 | 21 |

| Category | METHOD | | | | | MODEL LINEARITY | | | PAPER FOCUS | | Total |
|---|---|---|---|---|---|---|---|---|---|---|---|
| | Global SA | OAT SA | Global UA | OAT UA | Other/Unclear | Linear | Nonlinear | Unclear | Method | Model | reviewed |
| MATHS | 21 | 15 | 3 | 2 | 6 | 4 | 24 | 13 | 11 | 29 | 40 |
| MAT_SCI | 13 | 4 | 1 | 1 | 0 | 0 | 16 | 2 | 0 | 18 | 18 |
| MEDICINE | 26 | 30 | 25 | 4 | 13 | 2 | 24 | 37 | 2 | 62 | 64 |
| PHAR_TOX | 2 | 2 | 9 | 1 | 3 | 1 | 11 | 5 | 1 | 18 | 19 |
| PHYS_ASTRO | 13 | 9 | 4 | 0 | 0 | 1 | 20 | 2 | 2 | 21 | 23 |
| SOC_SCI | 10 | 5 | 0 | 4 | 2 | 1 | 14 | 5 | 6 | 15 | 21 |

*3.1 Model focus versus application focus*

In addition to considering the approach to SA, the papers reviewed were categorised into either "method focus" or "model focus" (see "paper focus" column in Table 1).

*Model-focused* papers are defined as those which focus on a model, and use sensitivity analysis as a tool to investigate uncertainty or other aspects of the model. The primary conclusions of the paper are therefore related to the model.

*Method-focused* papers are those that introduce sensitivity analysis methodology, and use a model as a case study to demonstrate the new approach. Conclusions are therefore focused on the performance of the method, and results relating to the model are of secondary interest.

| Table 2. Scores across all disciplines. | | |
|---|---|---|
| Paper focus | | |
| | Method | 10% |
| | Model | 90% |
| Model linearity | | |
| | Linear | 7% |
| | Nonlinear | 61% |
| UA scope | Unclear | 32% |
| | OAT | 28% |
| | Global | 72% |
| SA scope | OAT | 42% |
| | Global | 58% |

Table 2 shows that most papers are unsurprisingly focused on the application, i.e. on the model at hand, and not on the methods. Of the total of 280 papers, 35 were methodological, i.e. having SA methods as their subject. Of these, 24 advocate the use of global methods. On the one hand, this is encouraging because it shows that global methods are being promoted. On the other hand, a small but significant fraction of methodological papers are still advising statistically-incorrect OAT methods.

We note among the method papers a marked preference for variance-based measures of sensitivity – such as the sensitivity indices of which the Pearson correlation ratio discussed previously is a special case. Using these methods, one can decompose the variance of the output into terms of the first order (Pearson) and terms of higher orders describing interactions [9]. We also see an active line of research in moment-independent methods [34]. When using these latter, instead of looking at how factors affect the output variance, one looks at how they affect the empirical probability distribution of the output as generated by the UA. Precisely because they look at the entire distribution and not just a moment (such as the variance) they are called moment-independent.

*3.2 Model linearity*

If all models were linear, an OAT or derivative based approach would be adequate. This is because, in the linear case, the partial derivative of the output with respect to a given input is the same anywhere in the input space. Therefore, the partial derivative is a complete representation of sensitivity. In the nonlinear case, the partial derivative is *not* constant across the input space, so derivatives must be repeatedly sampled in different locations: this leads to global sensitivity analysis methods. In fact, the linearity or nonlinearity of the model is not always evident. Table 2 shows the proportions of linear and nonlinear models. Only in 8% of the cases were we able to conclude that the model was definitely linear, whereas over half of papers included clearly nonlinear models, with the remainder being unclear. This demonstrates that in the large majority of cases, global methods are essential to perform a methodologically-sound sensitivity analysis.

*3.3 Uncertainty Analysis*

Although, as discussed, uncertainty analysis and sensitivity analysis are distinct (but related) disciplines, in the literature the term "sensitivity analysis" is often used to describe both terms. As a result, the set of papers reviewed also included a number that were concerned with pure UA. These were not however excluded from the review, because OAT-based uncertainty analysis is as wrong as an OAT-based sensitivity analysis. Excluding UA would have resulted in missing an important part of the target audience of this work.

Of the 280 papers reviewed, 24 did not contain any kind of sensitivity analysis and instead only concerned uncertainty analysis: these represent clear conflations of sensitivity and uncertainty analysis.

One quite prevalent trend in some fields is the practice of performing a global UA (i.e. via a Monte Carlo analysis) side by side with an OAT SA: this was observed in particular, in Medicine, and Economics and Finance. In Medicine, for example, it seems to be common to perform an OAT sensitivity analysis, presenting the results in a tornado plot (a bar chart which shows the effect on the output of varying each assumption by a fixed amount in either direction). We speculate that the authors involved were not unaware of the chance to use elementary scatterplots of the output versus the input to rank the factors by importance – or simply they did not find this kind of analysis relevant or useful. In any case, once a certain practice becomes established within a given field (i.e. found in highly-cited papers), it sets a strong precedent which is difficult to supersede. Researchers and reviewers (not unreasonably) assume that if a method is found in influential articles then it must be correct.

Table 2 reports the occurrence of UA found in the literature review. In about ¾ of papers, there was either no UA present, or the methodology was not clearly specified. The former is due to the fact that our search query specifically targeted sensitivity analysis papers, so it is unsurprising that there are a large proportion of papers with little attention given to the UA part. On the other hand, about ¾ of the UAs that were observed *were* global in nature. This is most likely because a Monte Carlo analysis (randomly sampling from input distributions) is fairly intuitive and accessible to most researchers. Moreover, a so-called OAT uncertainty analysis (varying one input factor at a time and observing the maximum range of variation) is arguably less intuitive (as opposed to in sensitivity analysis where it is the opposite).

The same analysis can be applied by subject area: see Figure 6. Here we see that uncertainty analysis was found much more commonly in Pharmacology and Toxicology and Medicine (within the papers that we reviewed) than Social Sciences and Computer Science, for example. This should not be taken as an overall indication of the quantity of uncertainty analysis, because our sample has overwhelmingly targeted sensitivity analysis papers. However, it indicates that in Pharmacology and Toxicology and Medicine, either it is particularly common to perform UA simultaneously with SA, or the terms are confused. If we take the case of Pharmacology and Toxicology, we find that of the papers reviewed, only four had a sensitivity analysis, whereas ten had an uncertainty analysis. This flags that sensitivity analysis often refers to uncertainty analysis within this field.

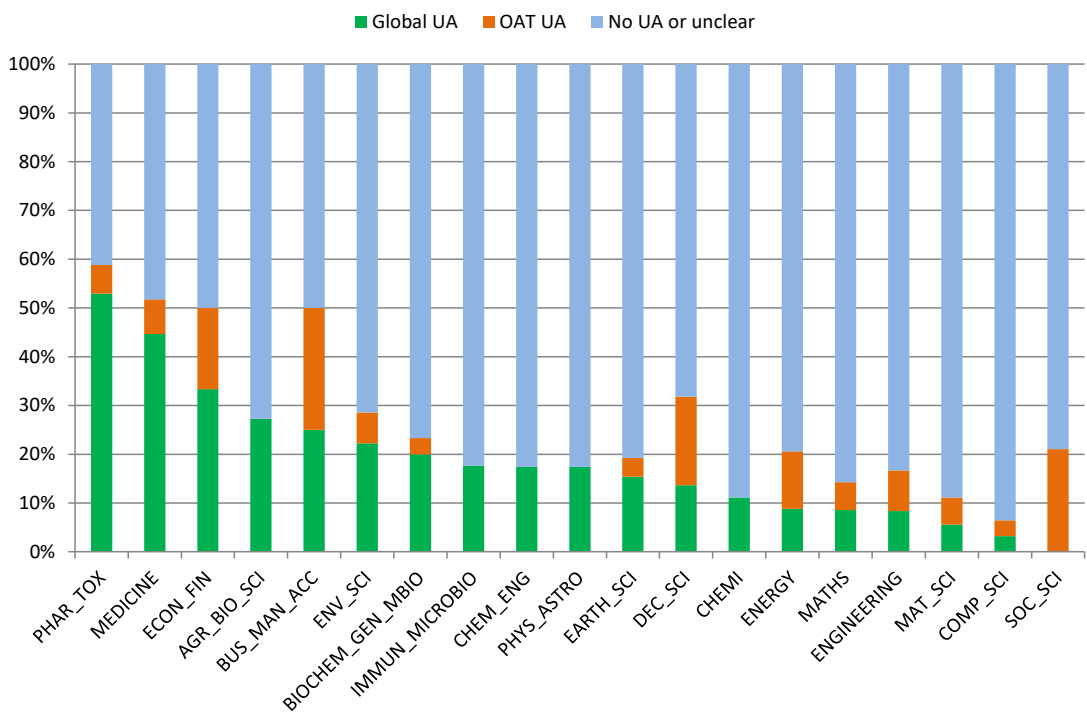

*Figure 6: Classification of uncertainty analysis by subject identifier, sorted by proportion of global methods*

*3.4 Sensitivity analysis*

Turning now to sensitivity analysis, Table 2 shows that 41% of sensitivity analyses use global methods, with 34% using OAT methods, and 25% having an unclear method type or no sensitivity analysis present. This is in slight contrast with a previous study which indicated a majority of sensitivity analysis to be OAT. This is encouraging, in that nearly half of studies use global methods. Still, around one third of highly-cited papers, matching our search criteria, use deficient OAT methods.

Figure 7 shows that the distribution of global methods varies widely across disciplines. Immunology and Microbiology show more than 70% of papers featuring global methods. This is followed by disciplines that are fairly model-intensive, such as Material Science, Biochemistry, Computer Science, and Engineering. At the other end of the spectrum, Pharmacology and Toxicology; and Business, Management and Accounting have very low proportions of global SA—about 10% and 20% respectively. Perhaps surprisingly, some disciplines that tend to rely heavily on large computer models, such as Earth Science and Environmental Science, still feature quite low rates of global sensitivity analysis. This is a concern, particularly when large-budget models are used for making significant decisions, such as climate models in policy-making—see a discussion in Saltelli et al. [35].

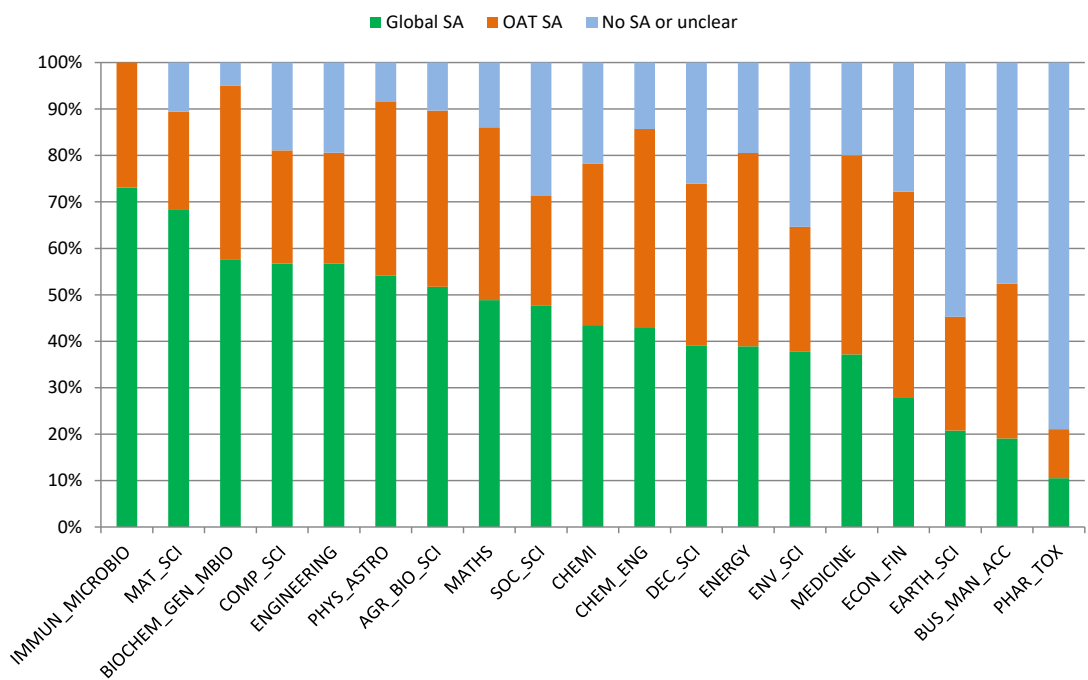

*Figure 7: Classification of sensitivity analysis by subject identifier, sorted by proportion of global methods*

## 3.5 A comparison with Ferretti et al.[36]

In a previous paper [36] we have investigated the prevalence of UA and SA methods based purely on the occurrence of keywords, i.e. without actually reading and reviewing the articles. The query to spot global SA papers relied on text mining, by identifying at least one known global sensitivity analysis technique (i.e. variance-based, metamodeling, elementary effects etc.). Figure 8 shows the results extended (present paper) to 2015 and 2016 (the original paper stopped at 2014). Here it would seem that only a modest fraction of papers that feature sensitivity analysis are global.

Two reasons explain the difference with the results in the present paper. First, as has been well-established here, "sensitivity analysis" is often also used to indicate uncertainty analysis, so that the upper curve in Figure 8 show a mixture of UA and SA, as well as an inevitable share of papers not pertained to mathematical modelling. Secondly, the estimation of the number of global SA papers is likely an underestimate because papers may apply simpler global methods, e.g. a scatterplot-based analysis, but not necessarily refer to the articles or techniques listed.

It is not among the aims of this paper to assess the evolution in the usage of SA and GSA in the global literature, as we deliberately chose to focus our analysis only on the most cited papers. Evidence (and indeed common sense) dictates that older papers tend to be cited more (see e.g. Davis and Cochran [37]), so it would not be meaningful to try to plot trends over time (almost half of the papers reviewed in this article are from 2012).

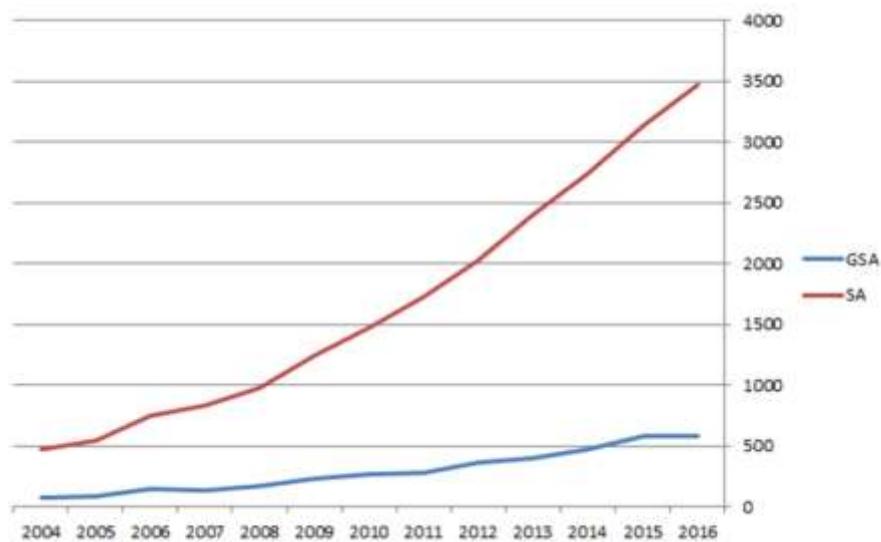

*Figure 8: Results from Ferretti et al., extended to 2016 (present paper)*

**Conclusions**

As discussed in the introduction, mathematical modelling is not a discipline per, and every branch of science and technology approaches modelling following its own culture and practice. One can also note that signals of distress as to the quality of mathematical modelling are heard from different disciplines: from economics[38,39] to natural sciences[10,40,41]. The situation has worrying analogies with what we have witnessed in data analysis, where misuse of the p-value has been singled out as one of the reasons of the present reproducibility crisis affecting science[24,30–32]. Our analysis points to likewise worrying signals for modelling arising from the poor quality of many sensitivity analysis. A considerable number of inaccurate papers have been published, i.e. papers where one-at-a-time methods have been used for either uncertainty or sensitivity analysis, in an environment where only a small proportion of models are demonstrably linear.

Determining the proportion of papers which use inadequate and inappropriate UA/SA methods is complicated by a lack of clarity about the linearity of the models being analysed.

Out of the 280 papers retained for the analysis 125 papers fall in this 'unclear' category for either method or model linearity. The data also indicate that there 57 papers in which either an uncertainty analysis or a sensitivity analysis was performed moving one factor at a time for a non-linear model.

Adopting the stance that this lack of clarity is, in itself, a significant methodological flaw, leads to the conclusion that 65% ((57+125)/280) of the reviewed papers are based on flawed methods. If only the papers where there was clarity about the linearity of the methods are considered then 37% (57/(280-125)) of papers are found to contain a fundamentally flawed approach to UA/SA. Even the most generous interpretation, where all papers with unclear linearity are given as good, results in over 20% (57/280) of papers being judged to contain inadequate and inappropriate UA/SA methods: a significant proportion and cause for concern.

OAT fails elementary considerations of experimental design and does not properly explore the space of the input factors. The result is that uncertainty is systematically under-estimated and sensitivity is wrongly estimated. Indeed, the sampling strategy of this meta-study deliberately targets highly-cited papers, under the assumption that they should represent at least average (and arguably better than average) practice in a given field, so the results here might even be seen as a slightly optimistic estimate of the real picture.

Though considerable differences exist in the use of sensitivity analysis among disciplines, all fields would benefit from the adoption of good practices. Our personal list of preferences, which agrees with the methodological papers seen in this review, would include the following recommendations:

- Both uncertainty and sensitivity analysis should be performed in general. Once an analyst has performed an uncertainty analysis and is informed of the robustness of the inference, it would appear natural to ascertain where volatility/uncertainty is coming from. At the other extreme, a sensitivity analysis without uncertainty analysis would also be illogical – the relative importance of a factor on the model output has a different relevance depending on whether the output has a small or large variance.
- Both uncertainty and sensitivity analysis should be based on an exploration of the input factors' space, be it using experimental design, Monte Carlo or other ad-hoc designs.
- When sensitivity analysis is performed, it should allow the relative importance of input factors to be assessed, either visually (scatterplots) or quantitatively (regression coefficients, sensitivity measures or other).

As regards what method should be used, our preference is for methods which are exploratory, model-free, capable to capture interactions and to treat group of factors (Box 1). A carefully performed uncertainty analysis, followed by sensitivity analysis, is an important ingredient of the quality assurance of a model as well as a necessary condition for any model-based analysis or inference. That such an analysis is still rare[36], and - when performed - often inaccurate (present work), demonstrate that action is urgent on the front of mathematical models' quality assurance procedures.